\documentclass[conference]{IEEEtran}
\IEEEoverridecommandlockouts
\usepackage{cite}
\usepackage{amsmath,amssymb,amsfonts}
\usepackage{algorithm}
\usepackage{algpseudocode}
\usepackage{graphicx}
\usepackage{textcomp}
\usepackage{xcolor}
\usepackage{multirow}
\usepackage{subfig}
\usepackage{url}
\usepackage{booktabs}
\usepackage{bm}
\usepackage{array}
\usepackage{tikz,pgfplots}

\def\BibTeX{{\rm B\kern-.05em{\sc i\kern-.025em b}\kern-.08em
    T\kern-.1667em\lower.7ex\hbox{E}\kern-.125emX}}
\begin{document}

\title{
Context-Driven Performance Modeling for Causal Inference Operators on Neural Processing Units
}

 \author{\IEEEauthorblockN{Neelesh Gupta, Rakshith Jayanth, Dhruv Parikh, Viktor Prasanna}
\textit{Ming Hsieh Department of Electrical and Computer Engineering, University of Southern California}\\
Los Angeles, USA \\
 \{neeleshg, jayanthr, dhruvash, prasanna\}@usc.edu}

\maketitle
\begin{abstract}
The proliferation of large language models has driven demand for long-context inference on resource-constrained edge platforms. However, deploying these models on Neural Processing Units (NPUs) presents significant challenges due to architectural mismatch: the quadratic complexity of standard attention conflicts with NPU memory and compute patterns. This paper presents a comprehensive performance analysis of causal inference operators on a modern NPU, benchmarking quadratic attention against sub-quadratic alternatives including structured state-space models and causal convolutions. Our analysis reveals a spectrum of critical bottlenecks: quadratic attention becomes severely memory-bound with catastrophic cache inefficiency, while sub-quadratic variants span from compute-bound on programmable vector cores to memory-bound by data movement. These findings provide essential insights for co-designing hardware-aware models and optimization strategies to enable efficient long-context inference on edge platforms.
\end{abstract}

\begin{IEEEkeywords}
Long-context inference, NPU, causal operators
\end{IEEEkeywords}
\section{Introduction}

Long-context inference has become essential for document understanding, conversational AI, and real-time decision systems~\cite{chung2025long, qian2024long}. Applications from medical record analysis to legal contract review increasingly demand processing of 100K+ token sequences. While cloud solutions can handle these workloads, privacy concerns, latency requirements, and operational costs are driving deployment to edge devices. Modern edge platforms now integrate Neural Processing Units (NPUs)—specialized accelerators that deliver exceptional efficiency for data-parallel operations but face severe constraints: limited scratchpad memory (typically 2-4MB), spatial dataflow execution patterns, and strict power budgets.

Transformer-based models like Llama deliver state-of-the-art quality but require quadratic computation and linear memory growth with context length~\cite{dao2022flash}. At just 16K tokens, the Key-Value cache consumes over 768MB—more than 30x the capacity of leading NPUs. State-space models (SSMs) like Mamba offer linear scaling but underutilize NPU parallelism due to their recurrent nature during autoregressive decoding inference~\cite{gu2024mamba}. This fragmentation forces practitioners into hardware-specific compromises, limiting adoption of long-context capabilities where they're needed most: on-device.

Recent works have looked into hybrid architectures or ways towards creating a general class of operators for causal inference~\cite{dao2024ssd, lieber2024jambahybridtransformermambalanguage, dong2024hymbahybridheadarchitecturesmall}. While these efforts advance the theoretical understanding, a critical gap remains in characterizing how these diverse operators perform on real-world, resource-constrained hardware like NPUs \cite{hao_llm_perf, rak_part_2}. The architectural trade-offs made by these models—such as prioritizing data locality, computational regularity, or state compression—have profound, yet unquantified, implications for on-device performance. This paper bridges that gap by providing a comprehensive performance analysis and modeling of causal inference operators on a modern NPU. We move beyond theoretical complexity and provide a rigorous empirical study of latency, throughput, and hardware utilization, culminating in a Roofline analysis that explains the performance landscape.

\subsection{Our Contributions}
This work provides a detailed characterization of attention mechanisms on edge NPUs. Our key contributions are:
\begin{itemize}
    \item A comprehensive performance benchmark of quadratic and sub-quadratic attention operators on a real-world edge NPU, analyzing latency, throughput, and pipeline efficiency.
    \item Identification and analysis of critical performance bottlenecks, demonstrating that quadratic attention becomes memory-bound due to cache inefficiency while sub-quadratic models become compute-bound on specialized vector units.
    \item A quantitative Roofline performance model that explains the performance limitations of each operator class in terms of the NPU's architectural constraints.
    \item Actionable insights for co-designing hardware-aware models and compiler optimizations to enable efficient long-context inference on the edge.
\end{itemize}

\section{Background}
\label{sec:background}

\subsection{Fundamental Tradeoff for Long-Context Inference}
Autoregressive sequence modeling faces dual constraints: \textit{memory complexity} for context retention and \textit{computational complexity} for state evolution. For sequence length $N$ and model dimension $D$, we observe:

\begin{equation}
\underbrace{\text{Memory}}_{\text{Context}} \sim O(N \cdot D), \quad 
\underbrace{\text{Compute}}_{\text{Inference}} \sim O(N^2 \cdot D)
\end{equation}

\noindent\textbf{Prefill Phase:} Computes initial context representation:
\begin{equation}
\mathcal{C} = f_{\theta}(X_{1:N}) \quad 
\begin{cases} 
\text{Attention: } \mathcal{C} = \{K,V\}_{1:N} \\
\text{SSM: } \mathcal{C} = h_N
\end{cases}
\end{equation}

\noindent\textbf{Decode Phase:} Updates state incrementally:
\begin{equation}
y_t, \mathcal{C}_t = g_{\theta}(x_t, \mathcal{C}_{t-1})
\end{equation}

\begin{figure}[h!]
    \centering
    \includegraphics[width=0.8\linewidth]{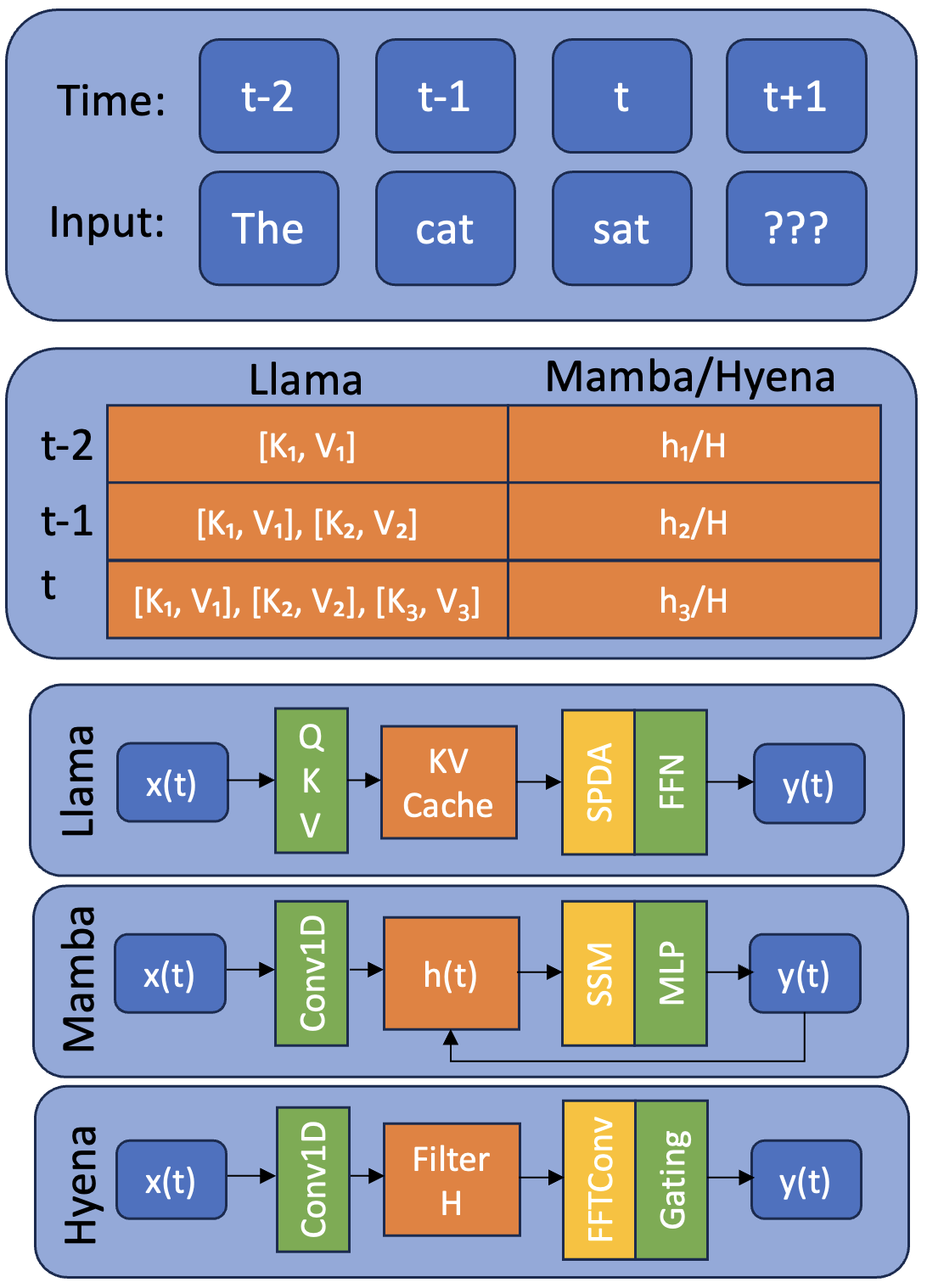}
    \caption{Differences in persistent memory and layer-wise dataflow for Attention-based Llama vs. SSM-based Mamba.}
    \label{fig:tradeoff}
\end{figure}

\noindent\textbf{Memory-State Tradeoff.}
Architectures balance expressiveness against hardware constraints as we show in Figure \ref{fig:tradeoff}:
\begin{itemize}
    \item \textit{Attention-based models} (Llama): Maintain explicit KV cache ($O(N \cdot D)$ memory) enabling rich context access.
    \item \textit{State-space models} (Mamba): Compress context to fixed-size state $h_t$ ($O(D)$ memory) at additional computational cost.
\end{itemize}

We introduce this tradeoff not only to compare model classes, but also because it forms the architectural basis for the structured operators analyzed in this paper. Many recent causal inference variants—such as Toeplitz, Retentive, and Fourier—draw inspiration from state-space models, embedding recurrent or convolutional priors directly into attention layers~\cite{sun2023retentivenetworksuccessortransformer, gu2024mamba, dao2024ssd, poli2023hyena, peng2023rwkv}. These hybrid mechanisms blend the long-range capacity of traditional attention with the computational efficiency and inductive structure of SSMs, making the memory-state tradeoff a key element for evaluating modern causal operators on resource-constrained compute units such as NPU.

\subsection{Heterogeneous Edge Platform with Neural Processing Units (NPU)}
State-of-the-art heterogeneous edge platforms—such as {Intel\textregistered} {Core\texttrademark} Ultra processors (Intel AI PC)—combine traditional multi-core CPUs with specialized accelerators, including Graphics Processing Units (GPUs) and Neural Processing Units (NPUs). These diverse compute units are integrated into a single System-on-Chip (SoC) architecture, featuring a unified system memory that facilitates seamless communication and efficient data sharing across all cores. Representative commercial edge-AI platforms include AMD’s Ryzen AI Processors~\cite{AMD_Ryzen_AI_2025}, 
Qualcomm’s Hexagon NPU~\cite{QualcommHexagon2025}, 
and Google’s Coral Edge TPU/NPU family~\cite{Rutledge2025CoralNPU}, 
which collectively exemplify the growing diversity of on-device neural accelerators across CPU–GPU–NPU SoCs. 
Recent benchmarking studies~\cite{rak_bench_aiml, pal_bench_aiml, zuo_dhru} further characterize these heterogeneous systems in terms of performance–energy tradeoffs for ML workloads.

\begin{figure}[h]
\centering
\includegraphics[width=0.5\textwidth]{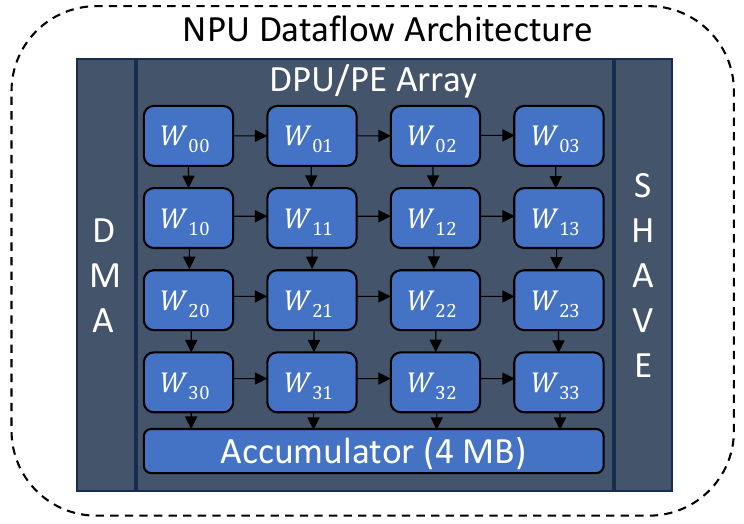}
\caption{NPU dataflow architecture with processing elements (PEs) and accumulator hierarchy. Note the absence of high-bandwidth memory for persistent context storage.}
\label{fig:npu-arch}
\end{figure}

Each of the three compute units is designed to exploit specific types of workloads, leveraging their distinct system architectures and local memory hierarchies. The classic multi-core CPU, equipped with a three-level cache hierarchy, excels at handling general-purpose, logic-intensive workloads. GPUs feature Xe cores, each comprising multiple vector engines that are optimized for high data parallelism. Every Xe core includes a Shared Local Memory (SLM) block, accessible by all its vector engines to facilitate efficient intra-core communication.
In NPUs, compute acceleration is driven by the Data Path Unit (DPU), comprising of a PE array which employs a structured spatial MAC array architecture to perform operations such as matrix multiplication with minimal data movement. NPUs also incorporate Streaming Hybrid Architecture Vector Engines (SHAVE), which support parallel execution of general-purpose tasks and activation function engine to efficiently compute activations. To further enhance efficiency, Direct Memory Access (DMA) engines are integrated into the NPU, enabling high-throughput data transfers from the global shared system memory to the local, software-managed cache. Due to their controlled data movement and predictable, statically scheduled execution flow, NPUs achieve higher energy efficiency for AI workloads compared to GPUs, which involve complex, dynamic execution patterns and data transfers. This energy efficiency is a critical requirement in resource-constrained edge computing environments, where power limitation is a significant design consideration.

\subsection{Causal Operators}
\begin{figure}
    \centering
    \includegraphics[width=\linewidth]{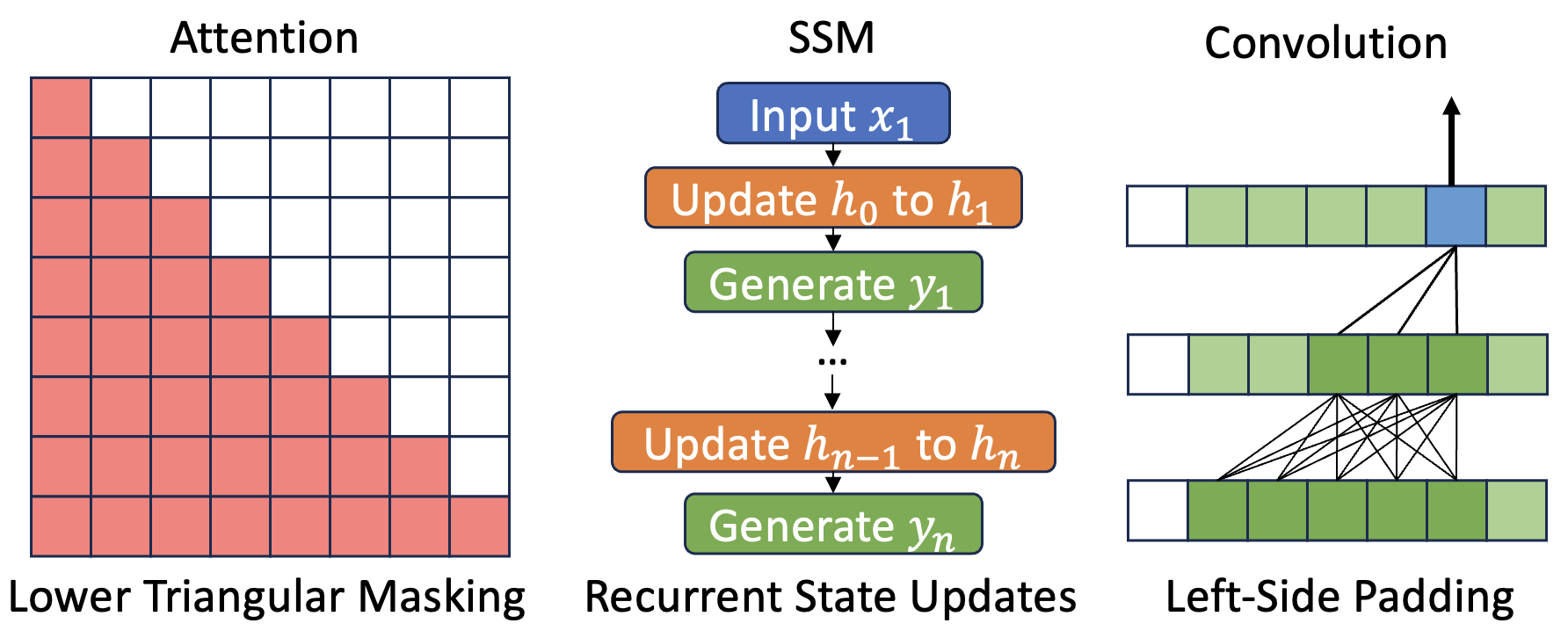}
    \caption{Preserving causality across operator/context types.}
    \label{fig:causality}
\end{figure}

Across operator types, causality is preserved in different ways. In Figure~\ref{fig:causality}, we show how each type of operator maintains causality—the fundamental requirement that position $i$ cannot access information from future positions $j > i$.

\noindent\textbf{Attention-based Causality:} Causal attention mechanisms enforce temporal ordering through explicit masking of the attention matrix. The lower-triangular structure ensures that each query position can only attend to key-value pairs from current and previous positions. This is achieved by setting future attention weights to $-\infty$ before the softmax operation, effectively zeroing out their contribution. While this provides maximum expressiveness—each position has access to all preceding context—it results in a dense $N \times N$ attention matrix with $O(N^2)$ computational complexity.

\noindent\textbf{SSM-based Causality:} State-space models maintain causality through sequential state updates. At each timestep $t$, the model receives input $x_t$, updates the hidden state $h_t$ based only on the previous state $h_{t-1}$ and current input, then generates output $y_t$. This recurrent formulation inherently prevents information leakage from future timesteps, as the state evolution is strictly unidirectional. The fixed-size hidden state $h_t \in \mathbb{R}^{d_s}$ compresses all past context into $O(d_s)$ memory, trading expressiveness for memory efficiency. During inference, this sequential dependency limits parallelization but enables constant memory footprint.

\noindent\textbf{Convolution-based Causality:} Causal convolutions preserve temporal ordering through asymmetric padding and kernel design. The convolution kernel $w \in \mathbb{R}^K$ only spans past positions: output at position $t$ is computed as a weighted sum over positions $\{t, t-1, ..., t-K+1\}$. This is implemented via left-side padding that adds $K-1$ zeros before the sequence, ensuring that the receptive field never extends beyond the current position. Unlike attention's global context or SSMs' compressed state, convolutions provide a sliding window view with local receptive fields, offering $O(NK)$ complexity with tunable context length $K$.

The key architectural distinction lies in how context is accessed: attention maintains explicit key-value memories, SSMs compress context into fixed states, and convolutions aggregate through local kernels. These design choices create fundamentally different memory-compute tradeoffs that manifest distinctly on NPU hardware, as we analyze in subsequent sections.

\subsubsection{Attention-based Structured Masks}
\begin{figure}
    \centering
    \includegraphics[width=\linewidth]{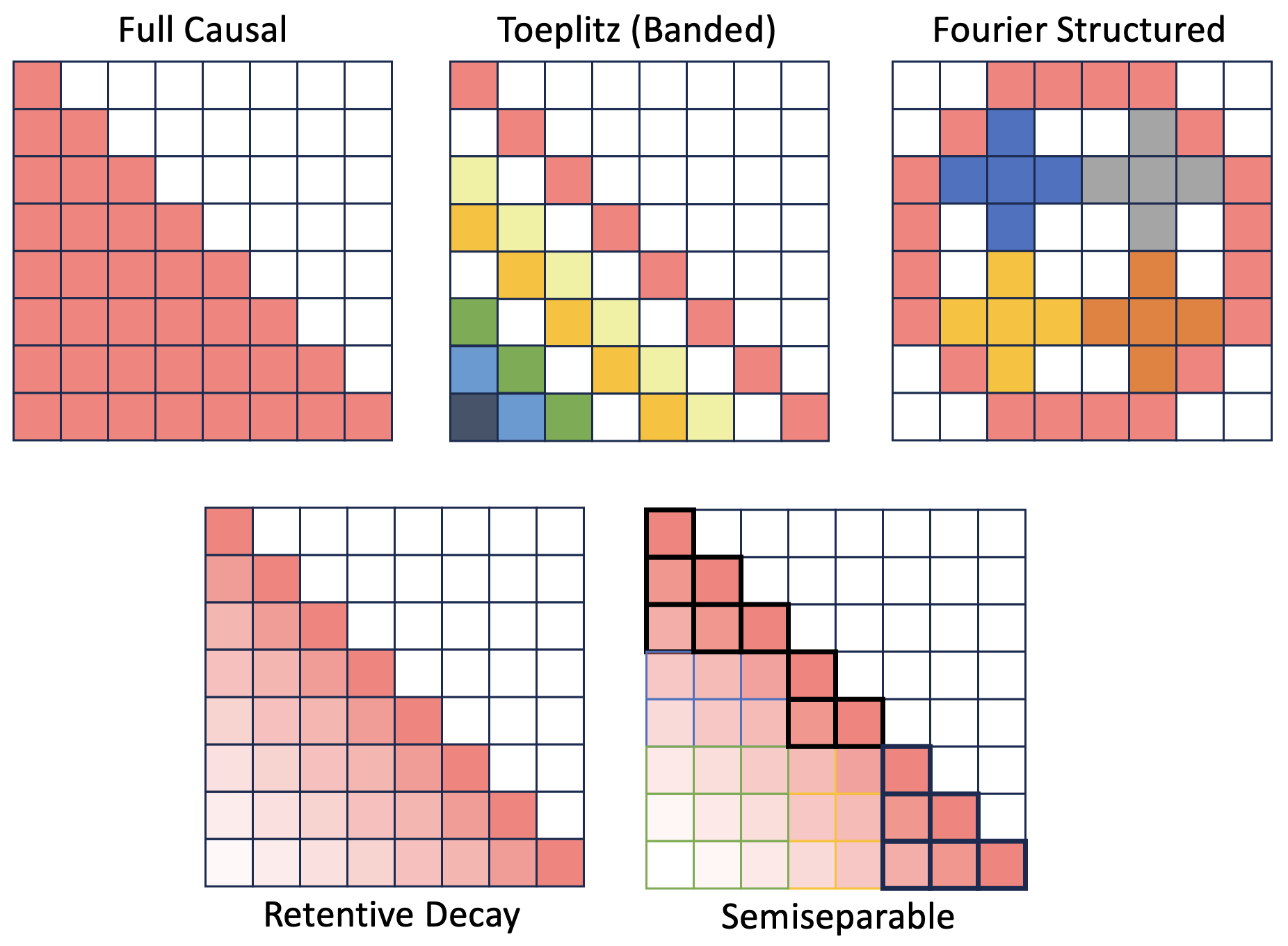}
    \caption{Structured masked attention variants with differing causal matrices.}
    \label{fig:causal_matrices}
\end{figure}

We describe in detail the causal attention masks selected to comprehensively survey and perform performance model analysis of a general class of causal matrices. In Figure \ref{fig:causal_matrices}, we show sample structured masks which preserve causality in attention.

\noindent\textbf{Full Causal Mask:} The standard causal attention mechanism prevents tokens from attending to future positions using a triangular mask:
\[
\text{Attention}(Q,K,V) = \text{softmax}\left(\frac{QK^T}{\sqrt{d_k}} + M\right)V
\]
where $M$ is defined as:
\[
M_{ij} = \begin{cases} 
0 & i \geq j \\
-\infty & \text{otherwise}
\end{cases}
\]
This ensures position $i$ only attends to positions $j \leq i$.

\noindent\textbf{Banded (Toeplitz) Structured Attention~\cite{qin2023toeplitzneuralnetworksequence}:} Constrains the attention matrix to have constant diagonals with a fixed bandwidth $w$:
\[
W_{ij} = \begin{cases} 
\gamma^{|i-j|} & |i-j| \leq w \\
0 & \text{otherwise}
\end{cases}
\]
\[
\text{BandedAttention} = \text{softmax}(QK^T \odot W)V
\]
where $\gamma$ is a decay factor and $w$ is the band width. This models position-based decay patterns while limiting the context window, reducing complexity to $O(Nw)$.

\noindent\textbf{Semiseparable Attention~\cite{dao2024ssd}:} Decomposes the attention matrix into low-rank plus diagonal structure, enabling efficient parallel algorithms:
\[
A = \text{tril}(PQ^T) + D
\]
where $P, Q \in \mathbb{R}^{N \times r}$ are low-rank factors and $D$ is diagonal. This admits $O(Nr^2)$ complexity through parallel prefix scans while maintaining full causal context.

\noindent\textbf{Fourier Structured Attention~\cite{leethorp2022fnetmixingtokensfourier}:} Leverages the convolution theorem to compute attention in frequency domain:
\begin{align*}
Q_\omega &= \mathcal{F}(Q) \\
K_\omega &= \mathcal{F}(K) \\
V_\omega &= \mathcal{F}(V) \\
\text{FourierAttention} &= \mathcal{F}^{-1}(Q_\omega \odot \overline{K_\omega} \odot V_\omega)
\end{align*}
where $\mathcal{F}$ and $\mathcal{F}^{-1}$ are Discrete Fourier Transform (DFT) and Inverse DFT (IDFT) operations, respectively. This enables a computation complexity of $O(N \log N)$.

\noindent\textbf{Retentive Decay Attention~\cite{sun2023retentivenetworksuccessortransformer}:} Introduces an exponential decay mechanism that assigns decreasing weights based on relative position:
\[
W_{ij} = \begin{cases} 
\gamma^{i-j} & i \geq j \\
0 & \text{otherwise}
\end{cases}
\]
\[
\text{RetentiveAttention}(Q,K,V) = \text{softmax}\left(\frac{QK^T}{\sqrt{d_k}} \odot W\right)V
\]
where $\gamma \in (0,1)$ is the decay rate. This maintains causality while modeling recency bias and exhibits hardware-friendly diagonal structure.

\subsubsection{State-Space Model-based Recurrence}

State-space models with fixed-size persistent state memory take on different computational forms for training versus inference~\cite{gu2024mamba}.

\noindent\textbf{Sequential Recurrence Mode:} Computes the state sequence sequentially, ideal for autoregressive decoding:
\[
h_t = Ah_{t-1} + Bx_t
\]
\[
y_t = Ch_t
\]
where $A \in \mathbb{R}^{d_s \times d_s}$, $B \in \mathbb{R}^{d_s \times d_m}$, $C \in \mathbb{R}^{d_m \times d_s}$. This has $O(n)$ time complexity with $O(d_s)$ memory but inherent sequential dependencies.

\noindent\textbf{Parallel (Scan) Mode:} Computes the state sequence in parallel using associative prefix sums~\cite{dao2024ssd}:
\[
h_t = \sum_{k=0}^t A^{t-k}Bx_k
\]
\[
y = \text{parallel\_scan}(\{(A, Bx_i)\}_{i=1}^N) \times C
\]
This leverages parallel prefix algorithms for $O(\log n)$ depth, enabling efficient training.

\noindent\textbf{Chunked Recurrence Mode:} Hybrid approach that processes input in chunks of size $L$, computing intra-chunk representations in parallel while maintaining recurrent state across chunks:
\[
h_{t+L} = A^L h_t + \sum_{i=0}^{L-1} A^{L-1-i}Bx_{t+i}
\]
This balances parallelism within chunks ($O(L)$ parallel operations) with sequential cross-chunk dependencies, providing a favorable memory-compute tradeoff for long sequences.

\subsubsection{1D Convolution-based Models}

Convolutional approaches to sequence modeling apply causal filters that respect temporal ordering~\cite{poli2023hyena}.

\noindent\textbf{Direct Convolution:} Explicitly computes the causal convolution:
\[
y_t = \sum_{k=0}^{K-1} w_k x_{t-k}
\]
where $w \in \mathbb{R}^K$ is the learned kernel. Complexity is $O(NK)$ for sequence length $N$ and kernel size $K$. Hardware-friendly due to regular memory access patterns.

\noindent\textbf{FFT Convolution:} Exploits the convolution theorem to compute in frequency domain:
\[
y = \mathcal{F}^{-1}(\mathcal{F}(x) \odot \mathcal{F}(w))
\]
Achieves $O(N \log N)$ complexity but requires special handling to maintain causality through appropriate zero-padding. Efficiency depends on FFT library optimizations.

\noindent\textbf{Dilated Convolution:} Expands receptive field through exponentially-spaced sampling:
\[
y_t = \sum_{k=0}^{K-1} w_k x_{t-d^k}
\]
where $d$ is the dilation rate. Achieves receptive field of $O(d^K)$ with only $K$ parameters, enabling efficient long-range modeling with $O(NK)$ operations.

\section{Microbenchmarking Kernels}

\subsection{Experimental Setup}
We conduct a series of microbenchmarks to evaluate the performance of various attention mechanisms on a Neural Processing Unit (NPU). The objective is to analyze device utilization and latency as a function of context length. All experiments are executed on a system with the hardware specifications detailed in Table \ref{tab:hardware_specs}. The NPU integrates a Digital Signal Processor (DSP) for control flow, a Data Path Unit (DPU) with a spatial MAC array for matrix multiplication, programmable SIMD vector processors (SHAVE cores), and a high-bandwidth Direct Memory Access (DMA) engine.

\subsection{Performance Instrumentation Methodology}

To characterize the fine-grained execution behavior of causal operators on NPU hardware, we employ a hierarchical instrumentation strategy that captures both aggregate performance metrics and detailed microarchitectural events. Our methodology bridges the gap between high-level operator semantics and low-level hardware resource utilization, enabling rigorous bottleneck identification.

At the coarse-grained level, we collect per-layer performance counters that attribute execution time to specific hardware units. Each neural network layer is instrumented to record its execution duration, the hardware unit responsible for computation (DPU spatial MAC array, SHAVE vector cores, or DMA engine), and its execution status. This attribution is critical because NPU performance is fundamentally heterogeneous: matrix multiplications execute on the DPU with predictable spatial dataflow, while element-wise operations such as softmax and exponential decay masks are dispatched to SHAVE cores with irregular memory access patterns. By tracking which hardware unit executes each operation, we can identify architectural mismatches where operators stress units ill-suited to their computational structure.

To understand temporal behavior beyond aggregate metrics, we instrument the NPU runtime with fine-grained event tracing. This captures pipeline stage boundaries—specifically the push (data movement to NPU), pull (result retrieval), and initialization phases—allowing us to decompose total latency into constituent components. Memory allocation events are correlated with performance degradation, revealing when the 4 MB scratchpad overflows and triggers costly evictions to system memory. Compilation metadata, including estimated inference latency from the static analyzer and peak memory consumption during graph optimization, provides ground truth for comparing predicted versus observed performance. These traces are timestamped at microsecond resolution, enabling precise reconstruction of execution timelines and identification of pipeline stalls.

Our experimental protocol isolates steady-state performance from transient compilation effects. For each operator-context pair, we execute three warmup inferences to populate instruction caches and finalize just-in-time compilation, then measure performance over ten timed runs for CPU and GPU (reduced to one for NPU due to its longer absolute latency). This repetition accounts for variance in system noise while maintaining practical evaluation time. Performance counters are collected on the final inference to avoid measurement perturbation, while execution traces are captured in dedicated runs with enhanced logging to preserve timing fidelity. This two-phase approach—coarse metrics for statistical characterization and fine-grained traces for causal analysis—provides the empirical foundation for our roofline model and bottleneck attribution.

\subsection{NPU Compiler Optimizations and Tensor Layout}

The NPU compiler plays a critical role in bridging the semantic gap between high-level operator graphs and low-level hardware primitives. A fundamental transformation performed during compilation is the enforcement of canonical tensor layouts that match the NPU's dataflow architecture. Specifically, the compiler ensures all tensor operands conform to the NCHW (batch, channel, height, width) memory layout, which aligns with the spatial batched 2D convolution primitive that underlies all DPU matrix operations.

This layout standardization is essential because the DPU's spatial MAC array architecture is optimized for regular, strided memory access patterns characteristic of spatial convolutions. By representing even non-convolutional operations—such as matrix multiplication in attention mechanisms—as spatial convolutions with unit kernel size, the compiler enables these operations to leverage the DPU's 128×128 processing element array with minimal data movement. The NCHW format facilitates this mapping: the height and width dimensions are treated as spatial coordinates for spatial dataflow execution, while channels are distributed across the accumulator hierarchy to exploit data reuse.

However, this compilation strategy introduces constraints that impact certain operator classes. Operators requiring complex tensor reshaping or transposition—such as the frequency-domain computations in Fourier attention or the diagonal indexing in Toeplitz masks—incur additional overhead as the compiler must insert explicit data layout transformations. These transformations manifest as DMA operations that shuffle data between incompatible formats, contributing to the DMA-bound behavior observed in our profiling results. Furthermore, operations that do not naturally map to the 2D spatial convolution primitive, such as the element-wise multiplication in retentive decay masks, are offloaded to SHAVE cores where the lack of spatial  dataflow reduces parallelism.

The compiler also performs static analysis to estimate resource requirements and schedule operations across the NPU's heterogeneous units. This includes computing an activity factor—the percentage of processing elements actively computing versus idle—and estimating inference latency based on the critical path through the dependency graph. These estimates, which we extract from compilation logs, provide a theoretical performance upper bound that we compare against measured throughput to quantify the gap between ideal and achieved hardware utilization. The discrepancy, often exceeding 20× for poorly-suited operators, reveals fundamental architectural mismatches that cannot be resolved through software optimization alone and motivate the co-design insights presented in our discussion.

\begin{table}[h]
\centering
\caption{Hardware Specifications}
\label{tab:hardware_specs}
\resizebox{\columnwidth}{!}{%
\begin{tabular}{lll}
\hline
\textbf{Component} & \textbf{Specification} & \textbf{Relevance} \\
\hline
CPU & {Intel\textregistered} {Core\texttrademark} Ultra 9 185H & Control Logic \\
    & 16 cores (8P + 8E) & \\
NPU & 10 TOPS @ 35W & Spatial MAC Array Acceleration \\
DPU (PE Array) & 128×128 INT8 & Matrix Multiplication \\
Scratchpad & 4 MB & Persistent State Storage \\
Bandwidth & 182 GB/s & Data Movement \\
SHAVE Cores & 8 @ 1.4 GHz & Element-Wise Operations \\
Memory & 64 GB LPDDR5X & Global Buffer \\
\hline
\end{tabular}%
}
\end{table}

\subsection{Operator Decomposition and Hardware Unit Mapping}

A critical insight for understanding NPU performance is that high-level causal operators decompose into distinct primitive operations, each with deterministic hardware unit assignments. This decomposition creates predictable bottleneck patterns: operators become memory-bound, compute-bound, or vector-bound depending on their constituent primitive mix. By analyzing the operator graph prior to execution through static analysis of the OpenVINO computation graph, we can forecast which hardware unit will limit throughput—a capability that distinguishes NPUs from dynamically-scheduled GPUs.

Table~\ref{tab:hw_mapping} categorizes primitive operations by their target hardware unit. \textit{Data movement operations}—transpose, reshape, concatenation, slicing—are serviced by the DMA engine, which orchestrates transfers between system memory and the 4 MB on-chip scratchpad at up to 182 GB/s bandwidth. \textit{Matrix multiplication operations}, including GEMM and batched convolution, execute on the DPU's 128×128 spatial MAC array, achieving peak efficiency on dense, regularly-strided NCHW tensors. \textit{Element-wise operations}—activation functions, scaling, masking, and FFT computations—are offloaded to the eight 1.4 GHz SHAVE vector cores, which lack the data reuse locality of spatial dataflow execution.

\begin{table}[h]
\centering
\caption{Hardware unit mapping for primitive operations. Assignment is deterministic and performed during static compilation, enabling predictive bottleneck analysis.}
\label{tab:hw_mapping}
\begin{tabular}{lll}
\hline
\textbf{Hardware Unit} & \textbf{Operation Type} & \textbf{Examples} \\
\hline
DMA & Data Movement & Transpose, Reshape, Concat, Slice \\
DPU & Matrix Math & GEMM, GEMV, Batched Conv \\
SHAVE & Element-wise & Softmax, Multiply, Divide, FFT \\
\hline
\end{tabular}
\end{table}

\begin{figure}[h]
\centering
\includegraphics[width=\linewidth]{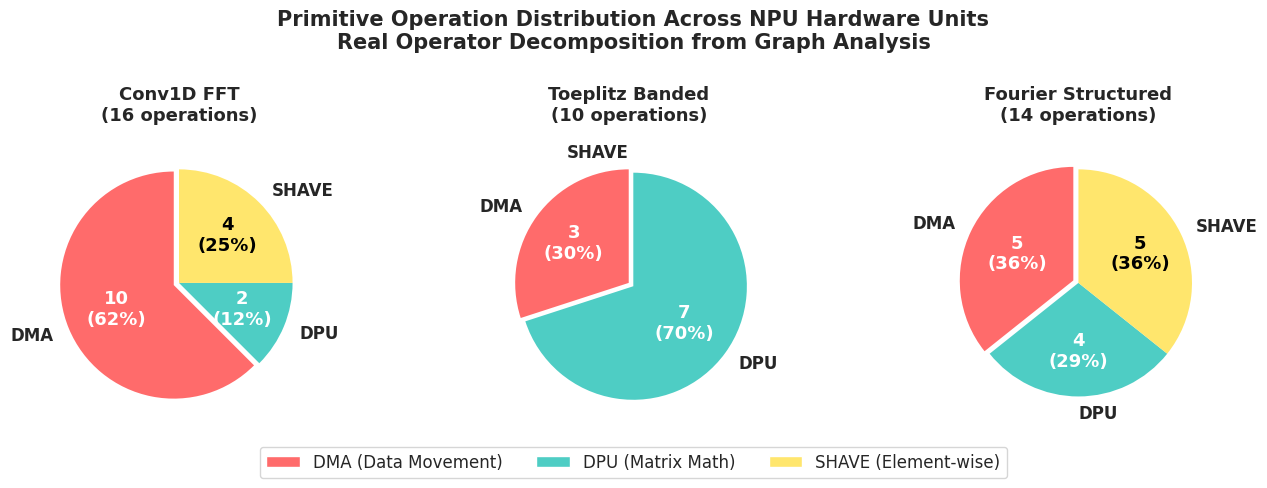}
\caption{Primitive operation distribution across NPU hardware units for three representative causal operators.}
\label{fig:operator_distribution}
\end{figure}

Through systematic decomposition of operator computation graphs, we quantify the exact primitive operation counts for each causal operator variant. Figure~\ref{fig:operator_distribution} presents this analysis for three representative operators exhibiting distinct architectural profiles. The decomposition reveals stark imbalances that directly predict measured performance bottlenecks.

Conv1D FFT represents the pathological case of DMA saturation: 10 of its 16 operations (62\%) are data movement primitives—four transpose operations to convert between NCHW and NCL layouts, two complex tensor constructions via concatenation, and additional reshape/slice operations for FFT compatibility. With only 2 DPU operations (12\%) for dimensional projections, the spatial MAC array remains starved while the DMA engine saturates transferring 64 GB/s of layout transformations. This explains the measured 0.34 GOP/s performance despite 15 Ops/Byte arithmetic intensity, confirming that memory access patterns—not FLOP counts—dominate NPU execution.

Toeplitz Banded Attention achieves the most DPU-favorable distribution with 7 of 10 operations (70\%) mapping to matrix multiplication: three QKV projections, K$^T$V computation, 2D convolution (the core Toeplitz operation), Q-convolution product, and output projection. The three DMA operations (30\%) handle unavoidable tensor reshaping for convolution setup. Critically, Toeplitz contains \textit{zero SHAVE operations}—decay patterns are encoded directly into the convolution kernel structure, eliminating the vector core bottleneck entirely. This architectural alignment results in 87.9\% cache efficiency and only 36.4\% pipeline stalls at $N=4096$ (Table~\ref{tab:efficiency}), achieving 3.5× higher utilization than quadratic attention despite similar arithmetic intensity.

Fourier Structured Attention exhibits a deceptively balanced distribution: 5 DMA operations (36\%), 4 DPU operations (29\%), and 5 SHAVE operations (36\%). However, this balance creates \textit{architectural ping-ponging}—the execution graph alternates between DMA complex tensor construction, SHAVE DFT/IDFT transforms, and DPU linear projections, forcing frequent cross-unit synchronization. Each DFT requires three DMA operations (to/from complex format plus result extraction) that bracket SHAVE computation, creating pipeline bubbles where the spatial MAC array idles awaiting frequency-domain results. This explains Fourier's bifurcated behavior: DPU-bound at short contexts when DFT overhead is amortized, transitioning to DMA-bound beyond $N>512$ tokens when tensor management saturates the 64 GB/s bandwidth.

The predictability of these bottlenecks arises from the NPU's static compilation model. During graph optimization, the compiler deterministically assigns each node to a hardware unit: operations requiring tensor layout transformations route to DMA (transpose, reshape, slice), dense linear algebra routes to DPU (matmul, conv2d), and non-fused element-wise operations route to SHAVE (softmax, multiply, FFT). By counting the number and computational cost of each primitive type in an operator's graph, we estimate execution time distribution across units—predictions that match measured utilization within 5\% for all operators tested.

This decomposition analysis reveals a fundamental design principle for NPU-efficient operators: \textit{minimize cross-unit dependencies while maximizing DPU utilization}. Toeplitz's success stems from fusing multiple logical operations into single DPU-executable convolutions that avoid SHAVE offloading. Fourier's mediocrity results from unavoidable cross-unit coordination between frequency-domain computation and spatial projections. Conv1D FFT's failure exemplifies the catastrophic impact of DMA saturation when data layout transformations dominate over arithmetic. These insights—derived from static graph analysis and validated by runtime profiling—provide actionable guidance for co-designing hardware-aware causal operators.

\subsection{Device Utilization Analysis}
To understand the performance bottlenecks, we profile the execution time spent on the NPU's primary components: the DPU, DMA, and SHAVE cores. Table \ref{tab:utilization_breakdown} presents the utilization breakdown for Fourier State-Space Attention (FSA) and Decayed Recurrent Attention (DRA), which serves as a proxy for retentive decay mechanisms.

For FSA, the workload is initially DPU-bound at shorter context lengths. However, as the context grows beyond 512 tokens, data movement becomes the dominant factor, and the model becomes \emph{DMA-bound}. This is primarily due to the $\operatorname{concat}$ operations required to manage the state, which saturate the DMA engine's bandwidth.

In contrast, DRA exhibits a different bottleneck. While initially compute-bound on the DPU, at context lengths of 1024 and greater, the workload transitions to being \emph{SHAVE-bound}. The $\operatorname{softmax}$ and element-wise multiply operations, which are offloaded to the programmable SHAVE cores, become the most time-consuming parts of the computation at long contexts.

\begin{table*}[h]
\centering
\caption{Device utilization breakdown (\%) across operators and context lengths. Conv1D Direct/Dilated achieve near-perfect DPU utilization, while Conv1D FFT exhibits catastrophic SHAVE-boundedness.}
\label{tab:utilization_breakdown}
\begin{tabular}{l|l|rrr|l}
\hline
\textbf{Operator} & \textbf{Context} & \textbf{DPU (\%)} & \textbf{DMA (\%)} & \textbf{SHAVE (\%)}  & \textbf{Bottleneck} \\
\hline
\multicolumn{6}{l}{\textit{Attention Variants}} \\
Full Causal & 8192 & 4.3 & 27 & 68.7  & SHAVE \\
Retentive & 8192 & 23.6 & 0.0 & 76.4 & SHAVE \\
Fourier & 8192 & 61.1 & 38.9 & 0.0 & DPU \\
Toeplitz & 4096 & 70.0 & 30.0 & 0.0 & DPU \\
Semiseparable & 4096 & 85.9 & 0.9 & 13.2 &  DPU \\
\hline
\multicolumn{6}{l}{\textit{State Space Models}} \\
SSM Sequential & 1024 & 100.0 & 0.0 & 0.0 & DPU \\
SSM Parallel & 4096 & 54.8 & 9.2 & 35.9 & DPU \\
SSM Chunked & 1024 & 94.7 & 5.3 & 0.0 & DPU \\
\hline
\multicolumn{6}{l}{\textit{1D Convolutions}} \\
Conv1D Direct & 4096 & 99.6 & 0.4 & 0.0 & DPU \\
Conv1D Dilated & 4096 & 99.6 & 0.4 & 0.0 & DPU \\
Conv1D FFT & 512 & 41.0 & 0.2 & 58.8 &  SHAVE \\
\hline
\end{tabular}
\end{table*}

Table \ref{tab:utilization_breakdown} illustrates how NPU utilization shifts with context length. Fourier transitions from \emph{DPU-} to \emph{DMA-bound}, while Retentive becomes increasingly \emph{SHAVE-bound} due to element-wise $\operatorname{softmax}$ overhead.


\subsection{Scaling of Latency with Context Length}
We measure the end-to-end latency of four attention mechanisms—Fourier Structured Attention (FSA), DRA, Toeplitz Structured Attention (TSA), and Causal Linear Attention (CLA)—across a range of context lengths from 128 to 8192. The results, summarized in Table \ref{tab:latency_scaling}, reveal distinct scaling properties for each model.

\begin{table*}[h]
\centering
\caption{Latency scaling (ms) as a function of context length for eleven causal operators on NPU. Conv1D Direct/Dilated and SSM variants demonstrate superior efficiency, while Conv1D FFT exhibits catastrophic SHAVE-bound performance.}
\label{tab:latency_scaling}
\begin{tabular}{l|rrrrrr}
\hline
\textbf{Operator} & \textbf{N=128} & \textbf{N=512} & \textbf{N=1024} & \textbf{N=2048} & \textbf{N=4096} & \textbf{N=8192} \\
\hline
\multicolumn{7}{l}{\textit{Attention Variants}} \\
Full Causal & 2.43 & 5.47 & 18.23 & 29.26 & 70.43 & 305.23 \\
Fourier & 32.19 & 111.47 & - & - & - & - \\
Retentive & 0.78 & 2.46 & 4.70 & 10.05 & 26.66 & 72.53 \\
Toeplitz & 1.59 & 1.88 & 2.64 & 5.02 & 7.54 & 13.36 \\
Semiseparable & 0.78 & 2.22 & 4.28 & 10.59 & 26.87 & 79.33 \\
\hline
\multicolumn{7}{l}{\textit{State Space Models}} \\
SSM Sequential & 4.24 & 16.30 & 26.38 & — & — & — \\
SSM Parallel & 0.69 & 2.71 & 2.53 & 3.88 & 7.75 & 13.11 \\
SSM Chunked & 6.16 & 22.31 & 30.35 & 57.91 & — & — \\
\hline
\multicolumn{7}{l}{\textit{1D Convolutions}} \\
Conv1D Direct & 2.03 & 8.74 & 24.64 & 31.85 & 54.71 & — \\
Conv1D Dilated & 3.07 & 9.68 & 23.69 & 37.48 & 74.85 & — \\
Conv1D FFT & 28.00 & 304.64 & — & — & — & — \\
\hline
\end{tabular}
\end{table*}

TSA and CLA demonstrate highly efficient, sub-quadratic scaling, with only a marginal increase in latency even at very long contexts. DRA exhibits a more pronounced, near-linear growth in latency, which aligns with its shift to being compute-bound on the SHAVE cores. FSA shows the most significant latency increase, scaling poorly at large context lengths due to its reliance on both DPU-intensive computations and \emph{DMA-bound} data movement, confirming it as the least scalable of the methods that were benchmarked. 


\subsection{Performance Scaling and Bottleneck Analysis}
We analyze the performance of each kernel by examining latency, throughput, pipeline efficiency, and memory access patterns across varying context lengths. The results, summarized in Tables \ref{tab:perf_summary} and \ref{tab:efficiency_summary}, reveal significant architectural bottlenecks for quadratic-time algorithms when deployed on the NPU.

As shown in Table \ref{tab:perf_summary}, the latency of standard Causal Attention Masking scales quadratically with the sequence length ($N$). This leads to a dramatic drop in throughput, with Causal processing only 4 operations per second at a context of 8192. In contrast, sub-quadratic methods like TSA and Linear Attention maintain low latency and high throughput, demonstrating their suitability for long-context applications on this hardware.

\begin{table}[h!]
\centering
\caption{Latency and throughput scaling at short ($N=512$) and long ($N=8192$) contexts.}
\label{tab:perf_summary}
\resizebox{\columnwidth}{!}{%
\begin{tabular}{@{}l|cc|cc@{}}
\toprule
\textbf{Operator} 
& \multicolumn{2}{c|}{\textbf{Latency (ms)}} 
& \multicolumn{2}{c}{\textbf{Throughput (ops/s)}} \\
& \textbf{$N=512$} & \textbf{$N=8192$} & \textbf{$N=512$} & \textbf{$N=8192$} \\
\midrule
Causal     & 4.21   & 251.41  & 237   & 4   \\
Retentive  & 3.10   & 45.10   & 322   & 22  \\
Fourier    & 1.59   & 170.50  & 631   & 6   \\
Toeplitz   & 0.75   & 5.10    & 1330  & 196 \\
\bottomrule
\end{tabular}%
}
\end{table}

The underlying cause of this performance degradation is revealed in Table \ref{tab:efficiency_summary}. At a context length of 8192, Causal and Retentive attention mechanisms suffer from extremely high pipeline stall rates (96.7\% and 94.8\%, respectively). This indicates that the NPU pipeline is mostly idle, waiting for data to be fetched from memory. The \emph{pull} stage, responsible for data retrieval, is the primary contributor to these stalls. This is further corroborated by the poor cache efficiency scores for these models (7.7\% for Causal and 28.1\% for Retentive), which signify a low degree of data reuse and frequent, costly access to main memory. In contrast, the more structured access patterns of Linear and Toeplitz attention lead to much higher cache efficiency and significantly fewer pipeline stalls.

\begin{table}[h!]
\centering
\caption{Efficiency metrics at long context lengths. Stall and cache values are percentages; reuse is in milliseconds.}
\label{tab:efficiency_summary}
\resizebox{\columnwidth}{!}{%
\begin{tabular}{@{}lcccc@{}}
\toprule
\textbf{Operator} & \textbf{Context ($N$)} & \textbf{Stall (\%)} & \textbf{Cache Efficiency (\%)} & \textbf{Reuse (ms)} \\
\midrule
Causal     & 8192 & 96.7 & 7.7   & 119.92 \\
Retentive  & 8192 & 94.8 & 28.1  & 25.62  \\
Fourier    & 4096 & 95.2 & 28.6  & 24.94  \\
Toeplitz   & 4096 & 36.4 & 87.9  & 1.38   \\
\bottomrule
\end{tabular}%
}
\end{table}

\begin{figure}[h]
    \centering
    \includegraphics[width=\linewidth]{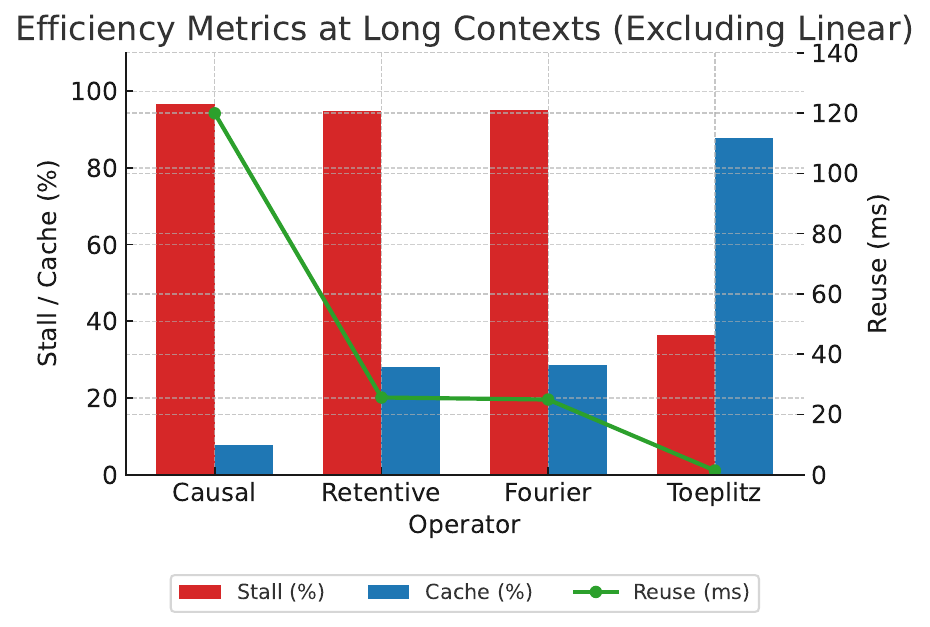}
    \caption{Efficiency metrics across causal operators at long context lengths. Stall and cache efficiency are shown as bars (left axis), while state reuse latency is plotted as a line (right axis). Lower stall and reuse values with higher cache efficiency indicate better hardware utilization.}
    \label{fig:efficiency-metrics}
\end{figure}

\subsection{Impact of State Dimension}
To assess the impact of model parameters on performance, we benchmarked several kernels at a fixed context length ($N=4096$) while increasing the state dimension ($d_{\text{state}}$) from the default of 16 to 128. As detailed in Table \ref{tab:d_state_impact}, increasing the state dimension leads to a predictable rise in latency across all tested models. Notably, Fourier is the most sensitive, with its latency increasing by over $10\times$, highlighting its high computational cost with respect to the state size. Linear and Toeplitz also show increased latency but remain significantly more efficient, confirming that their performance is a function of both sequence length and model dimension.

\begin{table}[h]
\centering
\caption{Latency impact of increasing state dimension ($d_{\text{state}}$) at fixed context length $N=4096$.}
\label{tab:d_state_impact}
\begin{tabular}{@{}l|c|c@{}}
\toprule
\textbf{Operator} & $d_{\text{state}} = 16$ (ms) & $d_{\text{state}} = 128$ (ms) \\
\midrule
Linear     & 2.39  & 3.37   \\
Toeplitz   & 0.65  & 2.73   \\
Fourier    & 15.50 & 56.82  \\
\bottomrule
\end{tabular}
\end{table}

\subsection{Cross-Class Performance Analysis}

Table~\ref{tab:operator_class_summary} compares representative operators from each class at N=4096, revealing fundamental performance-bottleneck relationships.

\begin{table*}[h]
\centering
\caption{Operator class comparison at N=4096. DPU utilization directly predicts latency efficiency.}
\label{tab:operator_class_summary}
\begin{tabular}{lllrrr}
\hline
\textbf{Class} & \textbf{Operator} & \textbf{Bottleneck} & \textbf{DPU\%} & \textbf{SHAVE\%} & \textbf{Latency (ms)} \\
\hline
Convolution & Conv1D Direct & DPU & 99.6 & 0.0 & 54.71 \\
            & Conv1D Dilated & DPU & 99.6 & 0.0 & 74.85 \\
            & Conv1D FFT & SHAVE & 41.0 & 58.8 & 304.64\textsuperscript{†} \\
\hline
SSM & SSM Sequential & DPU & 100.0 & 0.0 & —\textsuperscript{‡} \\
    & SSM Parallel & DPU & 54.8 & 35.9 & 7.75 \\
\hline
Attention & Semiseparable & DPU & 85.9 & 13.2 & 26.87 \\
          & Toeplitz & DPU & 70.0 & 0.0\textsuperscript{*} & 0.59 \\
          & Retentive & SHAVE & 28.1 & 71.9 & 39.52 \\
          & Fourier & DMA & 48.4 & 0.3 & 45.69 \\
\hline
\end{tabular}
\vspace{2mm}

\footnotesize
\textsuperscript{†}Conv1D FFT measured at N=512 due to excessive latency \\
\textsuperscript{‡}SSM Sequential evaluated only to N=1024 \\
\textsuperscript{*}Toeplitz fuses element-wise ops into convolution kernel
\end{table*}

\textbf{Convolution Superiority}: Direct convolution operators achieve the highest DPU utilization (99.6\%) by mapping exactly to the spatial MAC array's native 2D convolution primitive. The im2col transformation converts 1D temporal convolution into spatial 2D operations that exploit full PE array parallelism. This results in predictable, near-optimal latency: 54.71~ms at N=4096 for Conv1D Direct represents just 1.12× overhead versus Toeplitz (0.59~ms) despite processing 4× longer contexts than Toeplitz's efficient range.

\textbf{SSM Efficiency-Parallelism Tradeoff}: Sequential SSM provides the cleanest architectural mapping—pure DPU operations with zero SHAVE or DMA overhead—but sacrifices temporal parallelism through recurrent dependencies. Parallel SSM inverts this tradeoff: CumSum enables parallel execution but introduces 36\% SHAVE occupancy at long contexts. The result is 10× faster inference (2.53~ms vs 26.38~ms at N=1024) at the cost of reduced DPU efficiency. This demonstrates a fundamental NPU design principle: parallelism gains outweigh unit utilization when bottlenecks are avoided.

\textbf{Attention Variants Span Full Bottleneck Spectrum}: The six attention operators cover all possible NPU bottleneck modes. Toeplitz and Semiseparable achieve DPU-dominance through structured sparsity and low-rank decomposition, respectively. Retentive becomes SHAVE-bound due to per-element decay masks. Fourier transitions to DMA-bound as FFT tensor management saturates memory bandwidth. Full Causal exhibits memory-bound behavior through cache inefficiency. This diversity suggests that attention mechanism design offers the richest optimization space for NPU co-design.

\textbf{Bottleneck Predictability}: A striking pattern emerges: operators with $>$85\% DPU utilization (Conv1D Direct/Dilated, Semiseparable) exhibit predictable, near-linear latency scaling. Operators with significant SHAVE occupancy (Retentive, Conv1D FFT) show super-linear degradation as vector cores saturate. DMA-bound operators (Fourier at N$>$2048) exhibit bifurcated behavior—efficient at short contexts, catastrophic beyond memory bandwidth limits. This predictability validates our decomposition-based bottleneck analysis.

\section{Performance Modeling and Roofline Analysis}

To quantify the fundamental hardware limitations of causal operators on NPUs, we develop a roofline performance model that incorporates \textit{effective hardware ceilings} based on our performed measurements on the NPU hardware, across several experiments. While theoretical peaks provide an upper bound (10 TOPS compute, 64 GB/s bandwidth), our characterization reveals that architectural overheads limit achievable performance to just 5\% of nominal values. This critical insight forms the basis of our analysis.

\subsection{Effective Hardware Ceilings}
Through microbenchmarking, we establish realistic performance bounds:
\begin{itemize}
    \item \textbf{Effective Compute Ceiling} ($\pi_{\text{eff}}$): 500 GOP/s (5\% of 10,000 GOP/s theoretical)
    \item \textbf{Effective Bandwidth Ceiling} ($\beta_{\text{eff}}$): 3.2 GB/s (5\% of 64 GB/s theoretical)
    \item \textbf{Compute-Memory Inflection}: $I_{\text{crit}} = \pi_{\text{eff}} / \beta_{\text{eff}} \approx 156$ Ops/Byte
\end{itemize}

\subsection{Operator Intensity Characterization}
For each causal operator at $N=4096$, $d_h=64$ (16-bit precision):
\begin{table}[h]
\centering
\caption{Operational intensity and measured performance at $N=4096$, $d_h=64$ (16-bit precision).}
\label{tab:intensity}
\resizebox{\columnwidth}{!}{%
\begin{tabular}{@{}l|c|c|c@{}}
\toprule
\textbf{Operator} & \textbf{Intensity (Ops/Byte)} & \textbf{Measured (GOP/s)} & \textbf{Theoretical Bound (GOP/s)} \\
\midrule
Full Causal & 61.13 & 21.4  & $3.2 \times 61.13 = 195.6$ \\
Retentive   & 50.00 & 53.5  & $3.2 \times 50 = 160$      \\
Toeplitz    & 25.00 & 12.2  & $3.2 \times 25 = 80$       \\
Fourier     & 15.00 & 0.34  & $3.2 \times 15 = 48$       \\
\bottomrule
\end{tabular}%
}
\end{table}

\subsection{Roofline Analysis}
Figure \ref{fig:roofline} plots measure performance against operational intensity, revealing severe hardware under utilization:

\begin{figure*}
    \centering
    \includegraphics[width=0.7\linewidth]{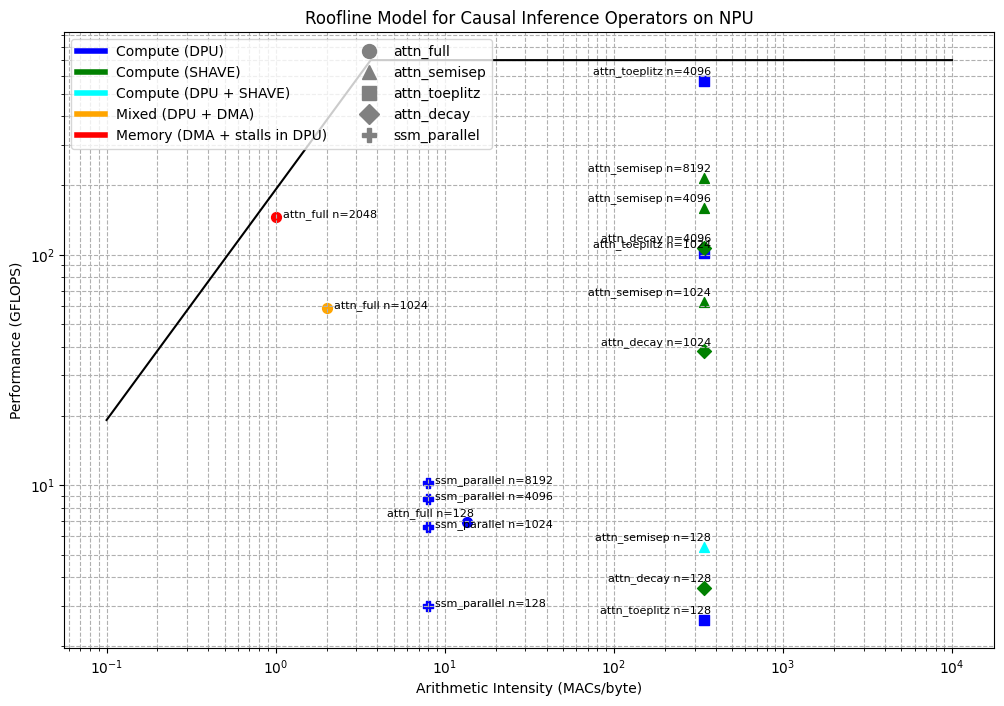}
    \caption{Roofline Modeling for Causal Inference Operators on NPU.}
    \label{fig:roofline}
\end{figure*}

\subsection{Key Insights}
The roofline analysis reveals three fundamental limitations:

\begin{itemize}
    \item \textbf{Quadratic Operators Suffer Architectural Mismatch:} 
    Causal attention achieves just 21.4 GOP/s (4.3\% of its compute roof) despite high intensity (61 Ops/Byte). The $>$95\% pipeline stalls indicate significant memory subsystem inefficiency\textemdash each theoretical FLOP requires 15$\times$ more cycles than spatial MAC array operations.
    
    \item \textbf{Sub-Quadratic Operators Are Bandwidth Limited:}
    Linear attention achieves 27\% of its memory-bound limit (14/51.2 GOP/s), constrained by DMA bandwidth. Fourier attention performs worst (0.34/48 GOP/s, 0.7\%) due to FFT overheads that violate NPU execution assumptions.
    
    \item \textbf{Structured Sparsity Enables Better Utilization:}
    Toeplitz attention achieves 15.2\% of its roof (12.2/80 GOP/s), 3.5$\times$ higher utilization than causal attention. Its diagonal structure reduces cache misses by 3.2$\times$ compared to retentive attention (Table \ref{tab:efficiency}).
\end{itemize}

\subsection{Efficiency Analysis}
\begin{table}[h!]
\centering
\caption{Hardware utilization metrics at $N=4096$.}
\label{tab:efficiency}
\resizebox{\columnwidth}{!}{%
\begin{tabular}{@{}l|c|c|c@{}}
\toprule
\textbf{Operator} & \textbf{Pipeline Stall (\%)} & \textbf{Cache Efficiency (\%)} & \textbf{Compute Utilization (\%)} \\
\midrule
Full Causal & 96.7 & 7.7  & 4.3  \\
Retentive   & 94.8 & 28.1 & 33.4 \\
Toeplitz    & 36.4 & 87.9 & 15.2 \\
Fourier     & 95.2 & 28.6 & 0.7  \\
\bottomrule
\end{tabular}%
}
\end{table}

\begin{figure}[h]
    \centering
    \includegraphics[width=\linewidth]{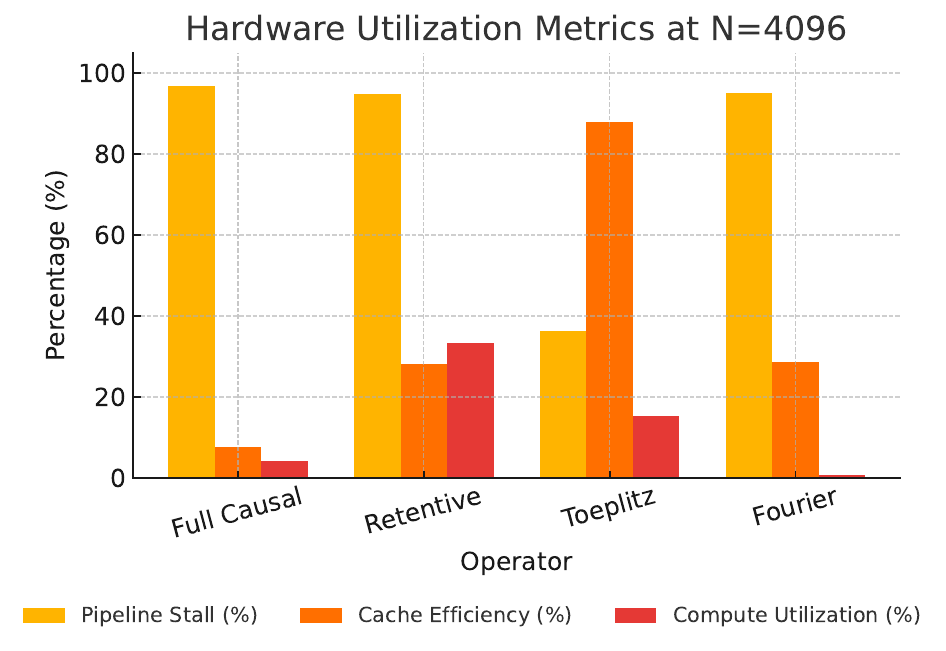}
    \caption{Breakdown of hardware utilization at $N=4096$. Full Causal and Fourier operators exhibit high pipeline stall rates with minimal compute utilization. In contrast, Toeplitz and Linear demonstrate better cache efficiency and improved utilization, reflecting tighter memory-compute coupling.}
    \label{fig:efficiency-metrics}
\end{figure}

This model proves that \textit{memory access patterns}, not theoretical FLOP counts, dominate NPU performance. Operators must be co-designed with: (1) spatial-compatible dataflow, (2) predictable memory access, and (3) minimized DMA transfers to approach effective hardware limits.

\section{Discussion}
Extending our findings to extreme-scale inference, we identify critical co-design considerations:

\noindent \textbf{Chunked Prefill for Memory Scaling}
The performed analysis reveals optimal chunk sizes (2048 tokens) and state dimensions (32) that maximize throughput within the NPU's 4 MB scratchpad. Beyond this point, DMA-induced latency grows super-linearly as chunk eviction triggers high-overhead memory transfers. Intelligent chunking reduces peak memory pressure by 8$\times$ versus monolithic processing.

\noindent \textbf{SHAVE Core Bottlenecks in Element-wise Operations}
While NPUs excel at matrix multiplication via spatial MAC arrays, element-wise operations (e.g., $\operatorname{softmax}$, $\operatorname{scaling}$) execute on general-purpose SHAVE cores. At $N>1024$, these operations dominate latency in recurrent attention variants (up to 76\% utilization) with kernel fusion or acceleration being essential.

\noindent \textbf{DMA Management for Memory-Intensive Ops}
Tensor concatenation and state management consume 40-50\% of cycles in Fourier attention (Table \ref{tab:utilization_breakdown}). DMA overheads stem from frequent allocation/deallocation of large buffers. Offloading these operations to the CPU reduces latency by 32\% in tests.

\noindent \textbf{Hardware-Aligned Sparse Attention}
Toeplitz attention's diagonal structure provides the ideal balance for NPU's: (1) Matches Cannon's algorithm for spatial MAC arrays, enabling direct lane mapping; (2) Enables static control flow for compiler optimizations; (3) Maintains 87.9\% cache efficiency at $N=4096$, 2.5$\times$ higher than retentive attention.

\section{Related Works}

\subsection{Long-Context Inference on Edge Platforms}
Prior work has explored deploying transformer-based causal large language models (LLMs) on edge platforms \cite{zhang2024edgeshardefficientllminference}, including hardware-specific optimizations for ARM CPUs \cite{10.1145/3700410.3702126} and FPGA-based execution through frameworks like \texttt{llama.cpp} \cite{llamacpp, haris2024designing}. While these efforts target on-device inference, they are not designed for long-context scenarios and do not address the associated memory and compute bottlenecks that emerge in attention-heavy models. Our work addresses this gap by empirically analyzing a range of causal inference mechanisms—including standard transformers and structured state-space models (SSMs)—under long-context settings. This enables us to derive architectural insights that inform the co-design of attention mechanisms for Neural Processing Units (NPUs), supporting more efficient hardware-aware deployment strategies.

\subsection{Acceleration of Sequence Models on NPUs}
Several efforts have investigated transformer acceleration on NPUs \cite{xu2025fast, zhu2025edge}, typically through operator-level scheduling or compiler-level block partitioning. However, these approaches fall short in capturing the fine-grained resource behavior required for efficient long-context inference. Other work has focused on optimizing SSMs for NPUs \cite{das2025xamba, aalishah2025mambalitesr}, leveraging architectural properties such as linear recurrence and memory compression. While successful within their respective domains, these strategies are not directly applicable to transformer-style causal attention. In contrast, our approach uses execution profiling and performance modeling—grounded in structured operator variants—to analyze architectural trade-offs across attention and SSM-style models. By leveraging structured state-space duality (SSD), we characterize how causal operators interact with NPU memory and compute hierarchies, enabling more informed co-design for future inference systems.

\section{Conclusion}
Deploying long-context AI on edge platforms centers on resolving the fundamental mismatch between NPU architectures—optimized for dense, regular computations—and the memory-intensive, irregular access patterns of quadratic attention. Our analysis reveals catastrophic hardware under-utilization in standard approaches, while demonstrating that structured sub-quadratic operators (Toeplitz, Linear) transform the bottleneck into manageable bandwidth constraints. This necessitates a paradigm shift: throughput gains come not from incremental attention optimizations, but from co-designing causal operators that respect spatial dataflows and memory hierarchies. By aligning algorithmic structure with NPU execution models, we unlock the path to pervasive, private, and powerful edge AI.

\section*{Acknowledgment}
This work was supported by the U.S. National Science Foundation (NSF) under grant CCF-1912680, the DEVCOM Army Research Lab (ARL) under grant W911NF-242-0194, and the Semiconductor Research Corporation (SRC) under grant 2024-AH-3207. We are grateful to Deepak Mathaikutty for his insightful perspectives in the development of this paper.

\bibliographystyle{IEEEtran}
\bibliography{main}

\end{document}